\documentclass[aps,prl,twocolumn,superscriptaddress, showpacs]{revtex4}
\usepackage{amsmath}
\usepackage{amssymb}
\usepackage{graphicx}
\usepackage{dcolumn}
\usepackage{bm}
\makeatletter
\def\captionof#1#2{{\def\@captype{#1}#2}}
\makeatother

\begin{document}

\title{Dynamical phase transition of a $1D$ transport process including death}
\author{S. Dorosz}
\affiliation{Department of Physics, Virginia Tech, Blacksburg, VA 24061, USA}

\author{S. Mukherjee}
\affiliation{Battelle Center for Mathematical Medicine, The Research Institute at Nationwide Children's Hospital, Columbus, OH 43205, USA}

\author{T. Platini}
\email{platini@vt.edu}
\affiliation{Department of Physics, Virginia Tech, Blacksburg, VA 24061, USA}


\begin{abstract}
Motivated by biological aspects related to fungus growth, we consider the competition of growth and corrosion. We study a modification of the totally asymmetric exclusion process, including the probabilities of injection $\alpha$ and death of the last particle $\delta$. The system presents a phase transition at $\delta_c(\alpha)$, where the average position of the last particle $\langle L \rangle$ grows as $\sqrt{t}$. For $\delta>\delta_c$, a non-equilibrium stationary state exists while for $\delta<\delta_c$ the asymptotic state presents a low density and max current phases. We discuss the scaling of the density and current profiles for parallel and sequential updates.
\end{abstract}

\date{\today}
\maketitle
Fungi are eukaryotic organisms that include micro-organisms such as yeasts and molds, as well as the familiar mushrooms. The growth of a fungus is formed by the combination of the apical growth and the branching process which leads to the development of a mycelium. Most of them grow as hyphae, which are a cylindrical thread-like structures of $5-10$ $\mu m$ in diameter and up to several centimeters in length \cite{Deacon}. The apical growth is a quasi one-dimensional process that extends the hypha by transport of material from the seed to the front tip. At this later position, enzymes are released into the environment, where the new wall material is synthesized. The rate of extension, in a favorable environment, can be extremely rapid, up to $40$ micrometers per minute. Many bio-physical models \cite{Prosser, Bezzi, Koch, Goriely,Gierz} describing the growth of fungal colonies and/or single hypha have been studied. In this context, efforts focused on the non-equilibrium properties of a modification of the totally asymmetric exclusion process (TASEP), by considering a distinct dynamics of one of the two boundary sites \cite{evans1,evans2,evans3,nowak,tailleur}.\\
An important aspect not taken into account yet, is that fungi have the ability to grow in a wide range of habitats, including extreme environments \cite{Vaupotic, Dadachova, Raghukumar} and survive intense UV/cosmic radiation during space travel \cite{Sancho}. Since the wall of the tip is usually structurally weak \cite{Deacon}, in such a situation the extension rate can be slowed down and as the hypha is progressively aging, it may break down or be broken by other organisms \cite{Deacon}. Our analysis is focused on the theoretical description of the above-mentioned growing process in competition with a corrosive environment. The theoretical model used is based on a simple modification of the TASEP and captures the general behavior induced by the two competing  processes.\\
Proposed in 1968 to study the motion of ribosomes along mRNA \cite{spitzer, MacDo}, numerous modifications of the TASEP were introduced including multilane systems and multi species transport \cite{karimi,adam,rein,ebbinghaus,greulich}, Langmuir dynamics \cite{frey,greulich2, santen}, extended particles \cite{dong2,antal} as well as systems with finite resources \cite{adams,cook}. In a general framework, such models are considered as toy models of transport phenomena in order to better understands physics far from equilibrium.

In the first part of this letter we define the model and the system parameters. After a careful definition of the parallel update dynamics we obtain, in the stationary state, the exact expression of the generating function and discuss the scaling of the density profile. In the last part, the results for the sequential update are presented. We are focusing our attention on the density and current profiles used to characterize the phase diagram. The system exhibits a dynamical phase transition between a region for which the size of the chain reaches a stationary value ($\langle L \rangle_{st}$) and a region in which the length is increasing with a constant velocity. At the transition line, the system exhibits a diffusive behavior $\langle L \rangle (t)\propto\sqrt{t}$. Finally we summarize our results and discuss areas for future work.
\begin{figure}
\centerline{\includegraphics[width=7.7cm,angle=0]{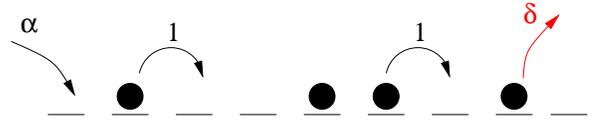}}
\caption{Sketch of the microscopic rules of the system. A particle is injected on the first site (if empty) with probability $\alpha$ while the last particle dies with probability $\delta$. In the bulk, particles are jumping forward with probability one if the site is empty.}
\label{fig0}
\end{figure}\\

Our model is a dynamical extended exclusion process (DEEP) described on a semi-infinite one-dimensional lattice. The dynamics is defined by the probability $\alpha$ of injection of a particle (on the first lattice site, if empty) and the probability $\delta$ of death of the last particle. In the bulk, the particles are hopping to the right with probability $p=1$, while the last particle (if not destroyed) jumps from site $L$ to $L+1$ with probability $1-\delta$ (see figure \ref{fig0}).  As a result of the microscopic rules, the length of the active chain $\langle L\rangle$ (defined by the rightmost occupied site) presents a dynamical phase transition. Extensions of this model can be considered including partially asymmetric diffusion and/or absorption/desorption processes of particles, so called Langmuir dynamics.

Considering first the case of the parallel update dynamics, we define the site occupation of the site $i$ at time $t$ by $\eta(i,t)$. The position of the last particle, at time $t$, is denoted by $L$ such that all sites $i>L$ are empty ($\eta(i,t)=0$). In one timestep, the last particle is updated first, it can jump forward or die. Then all the particles in the bulk are simultaneously moved forward and finally a particle is injected on the first lattice site with probability $\alpha$. Therefore the probability to have $n$ empty sites between two particles is given by $(1-\alpha)^n$. Knowing the configuration at time $t$, given by the set of $\eta(i,t)$ ($i\in {\mathbb N}^*$), the dynamical rules are:
\begin{eqnarray}
\eta(L+1,t+1)=1 &\text{with probability}& 1-\delta \notag \\
\eta(1<i\le L,t+1)=\eta(i-1,t) &\text{with probability}& 1 \notag \\
\eta(1,t+1)=1 &\text{with probability}& \alpha. \notag
\end{eqnarray}
After one timestep, the probability to find the last particle at the same position $L$ is given by the product $\alpha\delta$. This situation appears if we kill the last particle which is immediately followed by another one. By extension, knowing that $\eta(L,t)=1$, the probability to find, at time $t+1$, the last particle in $L-n$ is given by $\delta(1-\alpha)^{n}\alpha$. A transition is expected between two regions separated by the line $\delta_c(\alpha)$. For $\delta>\delta_c$, the last particle is killed at a sufficiently high frequency, that the growing process is effectively stopped. In this case a stationary state exists and we have $\langle L\rangle_{st}<\infty$. On the other hand, below the critical line the growth is only slowed down by the death of the last particle. The growing rate is finite, there is no stationary state and $\langle L\rangle(t)\propto t$. In the later situation we will analyze the asymptotic density and current profiles.\\
Starting in the region $\delta>\delta_c$, as we approach the critical line, the average $\langle L\rangle_{st}$ diverges. We can then neglect the effect of the left boundary and write the master equation for the probability $P(L,t)$ to find the last particle at site $L$ at time $t$. This describes a biased random walk with probability $1-\delta$ to jump to the right while the walker can jump to the left, over $n$ sites, with probability $\delta\alpha (1-\alpha)^n$. We can write explicitly
\begin{eqnarray}\label{EQA}
P(L,t+1)&=&(1-\delta)P(L-1,t)+\delta\alpha P(L,t)\notag\\
&+&\delta\alpha\sum_{k=1}^{\infty}(1-\alpha)^nP(L+k,t).
\end{eqnarray}
The time evolution of the average size is given by $\langle L\rangle(t+1)-\langle L\rangle(t)=1-\delta/\alpha$ and leads to the equation for the critical line $\delta_c=\alpha$. For $\delta>\delta_c$, an exact expression of the growing rate is given by the velocity $v=\langle L\rangle/t=1-\delta/\alpha$. At the critical point, the equation for the second moment leads to a diffusive law ($\langle L^2\rangle\propto t$) such that $\langle L\rangle=D_{\parallel} \sqrt{t}$. The coefficient $D_{\parallel}$ has been obtained from simulation data and evaluated to be (see figure $\ref{figDiff}$) 
\begin{eqnarray} \label{coefD}
D_{\parallel}=2\sqrt{{1-\alpha}}/\sqrt{{\pi\alpha}}+{\cal{O}}(10^{-3}).
\end{eqnarray}
The discrete time version of equation ($\ref{EQA}$), with the appropriate boundary equations (for $L=0,1$) can be cast into a single equation for its generating function defined by $G(z,t)=\sum_{L}z^LP(L,t)$. Analysis of the long time limit \cite{Zia} shows for $\delta<\delta_c$ that $\langle L\rangle \rightarrow (1-\delta/\alpha) t$ is recovered and leads, for $\delta > \delta_c$, to the exact stationary expression
\begin{eqnarray}
G^*(z)=\left(1-\frac{\alpha}{\delta}\right)\left(1+\frac{z\alpha}{1-\alpha-z(1-\delta)}\right).
\end{eqnarray}
Using $\langle L\rangle_{st}=\partial_{z} G^*(z)|_{z=1}$ we obtain the expression of the stationary length $\langle L \rangle_{st}=\alpha(1-\alpha)/\left[\delta(\delta-\alpha)\right]$. The order parameter $\langle L\rangle_{st}^{-1}$ vanishes linearly as we approach the critical line ($\langle L\rangle_{st}^{-1}\propto\epsilon$ with $\epsilon=\delta-\delta_c\ll1$).
Finally the stationary probability $P^*(L)$ (for $L\ne0$) is 
\begin{eqnarray}
P^*(L)=\left(1-\frac{\alpha}{\delta}\right)\frac{\alpha}{1-\delta}\left(\frac{1-\delta}{1-\alpha}\right)^L,
\end{eqnarray}
with $P^*(0)=1-\alpha/\delta$. The characteristic length $\lambda=[\ln((1-\alpha)/(1-\delta))]^{-1}$ diverges at the critical point.\\
The exact expression of the density profile, in the stationary state, is given by the sum $
\rho^*(n)=\alpha\sum_L \Theta(L-n)P(L)$, where $ \Theta$ is the Heaviside function. This leads to $\rho^*(n)=({\delta}/{\alpha})\langle N \rangle_{st} P(n)$, with $\langle N\rangle_{st}=\sum_{n\ge1}\rho^*(n)$ such that the average density $\bar{\rho}=\langle N\rangle_{st}/\langle L\rangle_{st}$ is equal to $\alpha$.\\
At the critical point, the scaling structure of the probability $P(L,t)$ is suggested by the diffusion law $\langle L\rangle\propto \sqrt{t}$ and leads to $P(L,t)=t^{-1/2}f(L/\sqrt{t})$. For large time, the asymptotic form is $f(u)\propto\exp(-u^2/\pi D_{\parallel}^2)$.  For $\delta<\delta_c$, the distribution is gaussian, centered in $\langle L\rangle(t)$ such that
\begin{eqnarray}
P(L,t)\propto\frac{1}{\sqrt{t}}\exp \left(-\left(\frac{L-\langle L\rangle(t)}{\sqrt{\langle L^2\rangle(t)-\langle L\rangle^2(t)}}\right)^2\right).
\end{eqnarray}
By integration we obtain the density profile given by
\begin{eqnarray}
\rho(x,t)=\alpha\left[ 1-{\rm{erf}}\left(\frac{x-\langle L\rangle(t)}{\sqrt{\langle L^2\rangle(t)-\langle L\rangle^2(t)}}\right)\right],
\end{eqnarray}
where ${\rm{erf}}(x)$ is the error function. The expression of the critical line could have been obtained from the equation $j=j_\delta$, where $j=\alpha$ is the particle current arriving at the interface while $j_\delta=\delta$ is the particle current leaving the system due to the death process.\\

\begin{figure}
\centerline{\includegraphics[width=7.5cm,angle=0]{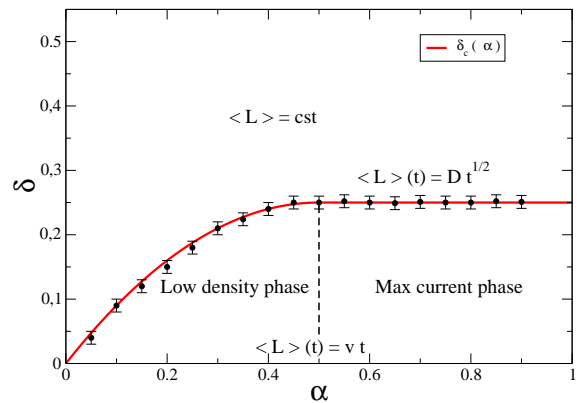}}
\caption{Phase diagram of the model. The critical line between the stationary and diverging regions is given by the equation $\delta_c=\alpha(1-\alpha)$ for $\alpha<1/2$ and $\delta_c=1/4$ for $\alpha>1/2$, the points are obtained via Monte Carlo simulations. The diverging region is divided in a low density and max current phases.}
\label{fig21}
\end{figure}

\begin{figure}
\centerline{\includegraphics[width=7.5cm,angle=0]{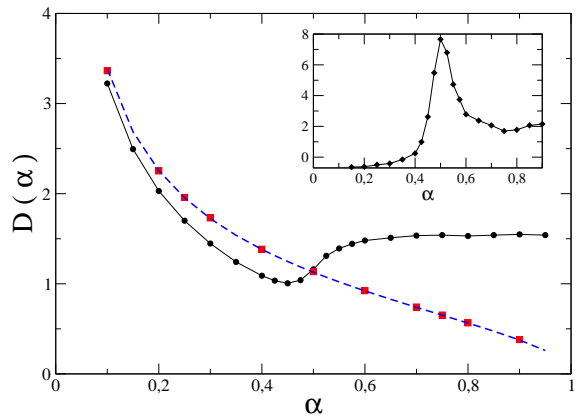}}
\caption{Coefficient of diffusion as a function of $\alpha$. The red squares are numerical results obtained for parallel updates and are in agreement with the expression ($\ref{coefD}$) - dashed line. The black circles, for sequential updates, present a non monotonic evolution. In the insert, the function $\partial_{\alpha} (D_{\perp}-D_{\parallel})$ presents a signature of the transition at $\alpha=1/2$.}
\label{figDiff}
\end{figure}

\begin{figure}
\centerline{\includegraphics[width=7.5cm,angle=0]{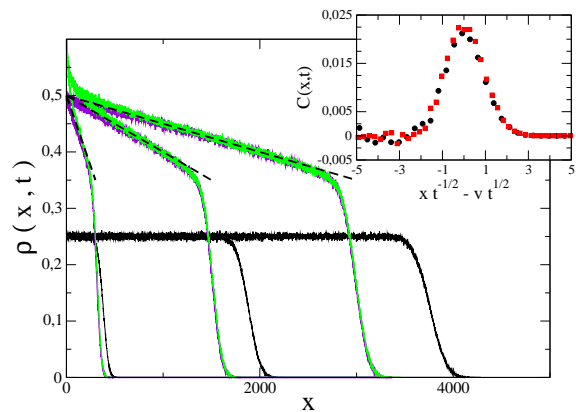}}
\caption{Density profile as a function of the position plotted for the times $t=1000,5000$, and $t=10000$. The black, violet and green curves are obtained respectively for the parameters ($\alpha=1/4$, $\delta=3/32$), ($\alpha=1/2$, $\delta=1/8$) and ($\alpha=3/4$, $\delta=1/8$). The dashed lines are given by the expression ($\ref{DD}$). In the insert, the correction $C(x,t)$ is plotted for the times ($t=5000$ and $t=10000$) and for $\alpha=1/2$, $\delta=1/8$.}
\label{figDD}
\end{figure}

\begin{figure}
\centerline{\includegraphics[width=7.5cm,angle=0]{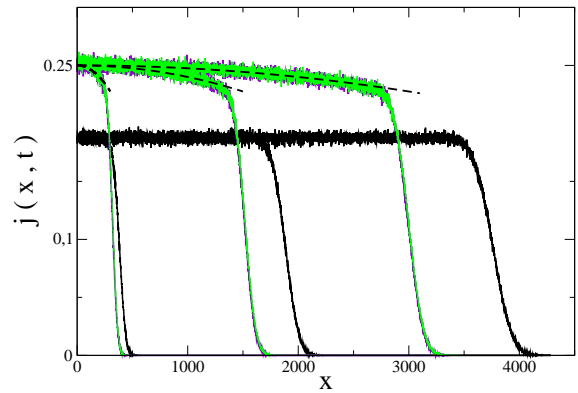}}
\caption{Current profile as a function of the position $x$ plotted for the times $t=1000,5000$, and $t=10000$. The black, violet and green curves are obtained respectively for the parameters ($\alpha=1/4$, $\delta=3/32$), ($\alpha=1/2$, $\delta=1/8$) and ($\alpha=3/4$, $\delta=1/8$). The dashed lines are given by the expression ($\ref{DD}$).}
\label{figJJ}
\end{figure}

For sequential update dynamics, at each timestep, ${\cal N}(t)+1$ updates are realized by choosing at random among the ${\cal N}(t)$ particles (present on the lattice at time $t$) and the first lattice site. Then on average, during one timestep each particle is updated once. When the rightmost particle is updated it can be removed from the lattice or moved forward. If a particle of the bulk is selected it moves systematically forward with respect to the exclusion process. Finally when the first lattice site is updated, a particle is injected with probability $\alpha$ if the site is empty. The analysis of the system behavior reveals a more complex phase diagram showed figure $\ref{fig21}$. In the region $\delta<\delta_c$, the chain size $\langle L\rangle$ is diverging and, for large time, the system properties should be similar to an infinite TASEP with effective $\beta_{eff}=1$. The critical line is then obtained when the current of particles arriving at the interface ($j$) is equal to the current of particles leaving the system ($j_\delta=\delta$). Since, for $\delta<\delta_c$, the system is expected to behave like the TASEP another transition line should exist for $\alpha=1/2$, between the low density and max current phases. The current of particles is given by $j=\alpha(1-\alpha)$ for $\alpha<1/2$ and $j=1/4$ for $\alpha\ge1/2$. We claim that the exact expression of the critical line is given by $\delta_c=\alpha(1-\alpha)$ for $\alpha<1/2$ and $\delta_c=1/4$ for $\alpha\ge1/2$. This is confirmed by the Monte Carlo simulations (fig. $\ref{fig21}$). In order to characterize these two phases, we are focusing our attention on the density and current profiles.\\
It is seen (figure \ref{figDD}) that the quadratic relation which usually expresses the current as a function of the local density of particles, $j=\rho(1-\rho)$, does not hold close to the interface ($x\simeq \langle L\rangle$) but is recovered in the bulk. The correction $C(x,t)$ defined as $\rho(x,t)(1-\rho(x,t))-j(x,t)$ is shown as an insert of figure \ref{figDD} for $\alpha=1/2$ and $\delta=1/8$. For large time, $C(x,t)$ is gaussian and rescaled as $C(xt^{-1/2}-vt^{1/2},t)$, where $v$ is the front velocity.\\
In the low density phase, the density is constant given by $\rho=\alpha$, while in the max current phase, as observed in \cite{evans2}, the current and density profiles continue to evolve as the system size is growing. The asymptotic expressions of the profiles are obtained assuming $x/t^z$ being the scaling variable of the density profile. Using the mean field relation $j(i,t)=\rho(i,t)(1-\rho(i+1,t))$, one obtains in the long time and continuous space limit $j(x,t)=\rho(x/t^z)(1-\rho(x/t^z))+{\cal O}(t^{-z})$. Together with the continuity equation $\partial_t\rho+\partial_xj=0$, the density is given by $\rho(x/t^z)=\frac{1}{2}\left(1-z\frac{x}{t}\right)$ which imposes $z=1$. Finally, for large times, the bulk density and current profiles are
\begin{eqnarray}\label{DD}
\rho(x/t)=\frac{1}{2}\left(1-\frac{x}{t}\right),\hphantom{AA}j(x/t)=\frac{1}{4}\left(1-\frac{x^2}{t^2}\right).
\end{eqnarray}
The numerical data and analytic predictions are in perfect agreement as shown in figures $\ref{figDD}$ and $\ref{figJJ}$.\\ 
In the stationary region, the probability distribution $P^*(L)$ as well as the density and current profiles decay exponentially, while on the critical line the decay of the density and current profiles is gaussian. Interestingly, for the sequential update dynamics, the coefficient of diffusion ($D_{\perp}=\langle L\rangle/\sqrt{t}$) presents a non monotonic evolution as a function of $\alpha$ (figure $\ref{figDiff}$). In the low density phase, since the number of interactions between particles is small, the same behavior is observed for $D_{\parallel}$ and $D_{\perp}$. We numerically checked that the first correction is of order $\sqrt{\alpha}$ such that $D_{\perp}\simeq D_{\parallel}+{\cal O}(\sqrt{\alpha})$. However, while $\alpha$ increases, the interactions between particles occur more often and  lead to a transition from the low density phase to the max current phase. A signature of this transition is found in the difference $D_{\perp}-D_{\parallel}$. In the insert of figure $\ref{figDiff}$, we plot the derivative $\partial_{\alpha}(D_{\perp}-D_{\parallel})$ which presents a maximum at the transition line $\alpha=1/2$.

In summary, motivated by the competition between growth and corrosion processes, we obtained the phase diagram of a dynamical extended toy model inspired by mycology problems. For both parallel and sequential updates, the model presents a dynamical phase transition between a region of finite size and a diverging region. For the sequential update, the transition between the low density and max current phases leaves a signature in the difference $D_{\perp}-D_{\parallel}$. It is also seen that the quadratic law $j=\rho(1-\rho)$ fails close to the interface but remains valid in the bulk. As observed in \cite{evans2}, in the max current phase, the density and current profiles are time dependent. In this region, we proved that the profiles are function of the scaling variable $x/t$ and independent of the boundary conditions. An identical dynamical phase transition, characterized by the same diffusive behavior on the transition line, was observed in \cite{Antal-Redner} on microtubule dynamics for which the growth enters in competition with a detachment process. It suggests that universal features might emerge close to the transition line generated by the competing processes. We motivate the experimental characterization of this transition by the analysis of fungus growth under UV radiation used as a control parameter.

We would like to thank the group of statistical mechanics of Virginia Tech for its support and particularly J. Cook, Professors Zia, Pleimling, Kulkarni, D. Karevski and R.J. Harris for useful discussions. This research is funded in part by the US National Science Foundation through DMR-0705152 and DMR-0904999.


\begin{thebibliography}{99}

\bibitem{Deacon} J, W. Deacon, in {\it Introduction to modern Mycology}, published by Wiley, John \& Sons (1997).

\bibitem{Prosser} J. Prosser, in {\it The  Growing Fungus}, edited by N. A. R. Gow and G. M. Gadd (Chapman and Hall, London, 1995), pp. 319-333.

\bibitem{Bezzi} M. Bezzi and A. Ciliberto, Comm. Theor. Bio. {\bf 8}, 585 (2003).

\bibitem{Koch} A. L. Koch,
Adv. Microb. Physiol. {\bf 29}, 301 (1983).

\bibitem{Goriely} A. Goriely and M. Tabor,
J. Theor. Biol. {\bf 222}, 221 (2003).

\bibitem{Gierz} G. Gierz and S. Bartnicki-Garcia,
J. Theor. Biol. {\bf 208}, 151 (2001).

\bibitem{evans1}
M. R. Evans and K. E. P. Sugden,
Physica A {\bf 384}, 53 (2007).

\bibitem{evans2}K. E. P. Sugden, M. R. Evans,
J. Stat. Mech., P11013 (2007).

\bibitem{evans3} K. E. P. Sugden et al.,
Phys. Rev. E {\bf 75}, 031909 (2007).

\bibitem{nowak}S. A. Nowak, P. Fok, and  T. Chou,
Phys. Rev. E. {\bf 76}, 031135 (2007).

\bibitem{tailleur}J. Tailleur , M. R. Evans and Y. Kafri,
Phys. Rev. Lett. {\bf 102}, 118109 (2009).

\bibitem{Vaupotic} T. Vaupotic et al.,
Fungal Genetics and Biology {\bf 45} (6) 994Ð1007 (2008).

\bibitem{Dadachova} E. Dadachova et al.,
PLoS ONE {\bf 2} e457 (2007).

\bibitem{Raghukumar} C. Raghukumar and S. Raghukumar,
Aquatic Microbial Ecology {\bf 15}, 153Ð63 (1998).

\bibitem{Sancho} L.G. Sancho et al.,
Astrobiology {\bf 7} (3) 443Ð54 (2007).

\bibitem{spitzer}F. Spitzer, 
Adv. Math. {\bf 5}, 246 (1970).

\bibitem{MacDo} C. T. MacDonald, J. H. Gibbs, and A. C. Pipkin, 
Biopolymers {\bf 6}, 1 (1968).

\bibitem{karimi}V. Karimipour, 
Europhys. Lett. {\bf 47}, 304-310 (1999).

\bibitem{adam}A. Brzank and G. M. Sch\"{u}tz, 
J. Stat. Mech., P08028 (2007).

\bibitem{rein}S. Klumpp and R. Lipowsky, 
Europhys. Lett. {\bf 66}, 90-96 (2004).

\bibitem{ebbinghaus} M. Ebbinghaus, L. Santen, 
J. Stat. Mech., P03030 (2009).

\bibitem{greulich}D. Chowdhury et al., 
European Physical Journal B {\bf 64}, 593 (2008).

\bibitem{frey} A. Parmeggiani, T. Franosch, and E. Frey, 
Phys. Rev. Lett. {\bf 90}, 086601 (2003).

\bibitem{greulich2}P. Greulich and A. Schadschneider, 
Phys. Rev. E {\bf 79}, 031107 (2009).

\bibitem{santen}M. R. Evans, R. Juhsz, and L. Santen, 
Phys. Rev. E {\bf 68}, 026117 (2003).

\bibitem{dong2}J. J. Dong, B. Schmittmann, and R. K. P. Zia, 
Phys. Rev. E {\bf 76}, 051113 (2007).


\bibitem{antal}T. Antal et al., 
J. Stat. Mech., P08027 (2007).

\bibitem{adams}D. A. Adams et al.,
J. Stat. Mech., P06009 (2008).

\bibitem{cook}L. J. Cook, R. K. P. Zia, and B. Schmittmann,
Phys. Rev. E {\bf 80}, 031142 (2009).

\bibitem{Zia} R.K.P. Zia, private communication, to be published.

\bibitem{Antal-Redner} T. Antal et al.,
J. Stat. Mech., L05004 (2007) 



\end{thebibliography}
\end{document}